\documentclass[10pt]{article}
\usepackage{algorithm,algorithmic,multirow,tabularx, graphicx, amsmath}

\begin{document}

\title{Turtle: Identifying frequent $k$-mers with cache-efficient algorithms}
\author {Rajat Shuvro Roy\,$^{1,2,3}\footnote{to whom correspondence should be addressed}$, Debashish Bhattacharya\,$^{2,3}$ and Alexander Schliep\,$^{1,4}$ \\
$^{1}$Department of Computer Science,  \\
$^2$Department of Ecology, Evolution and Natural Resources, \\
$^3$Institute of Marine and Coastal Sciences ,\\
$^4$BioMaPS Institute for Quantitative Biology.\\
Rutgers University, New Brunswick, NJ 08901, USA.}

\maketitle
\begin{abstract}
Counting the frequencies of $k$-mers in read libraries is often a
first step in the analysis of high-throughput sequencing
experiments. Infrequent $k$-mers are assumed to be a result of sequencing
errors. The frequent $k$-mers constitute a reduced but error-free
representation of the experiment, which can inform read error
correction or serve as the input to de novo assembly methods. Ideally,
the memory requirement for counting should be linear in the number of frequent
$k$-mers and not in the, typically much larger, total number of
$k$-mers in the read library.

We present a novel method that balances time, space and accuracy
requirements to efficiently extract frequent $k$-mers even for high
coverage libraries and large genomes such as human. Our method is
designed to minimize cache-misses in a cache-efficient manner by using
a Pattern-blocked Bloom filter to remove infrequent $k$-mers from
consideration in combination with a novel sort-and-compact
scheme, instead of a Hash, for the actual counting. While this
increases theoretical complexity, the savings in cache misses reduce
the empirical running times. A variant can resort to a counting Bloom
filter for even larger savings in memory at the expense of false
negatives in addition to the false positives common to
all Bloom filter based approaches. A comparison to the
state-of-the-art shows reduced memory requirements and running times.
Note that we also provide the first competitive method to count $k$-mers up to size 64.

The tools are freely available for download at \\{\tt http://bioinformatics.rutgers.edu/Software/Turtle}.\\
Contact: {\tt rajatroy@cs.rutgers.edu}

\end{abstract}

\section{Introduction}
{\em $K$}-mers play an important role in many methods in bioinformatics as they are at the core of the {\it de Bruijn} graph structure (\cite{Pevzner2001}) that underlies many of today's popular {\it de novo} assemblers (\cite{Simpson2009, Zerbino2008}). They are also used in assemblers based on the overlap-layout-consensus paradigm like Celera (\cite{Miller2008}) and Arachne (\cite{Jaffe2003}) as seeds to find overlap between reads. Several read correction tools (\cite{Kelley2010, Liu2012, Medvedev2011a}) use $k$-mer frequencies for error correction. Their main motivation for counting $k$-mers is to filter out or correct sequencing errors by relying on $k$-mers which appear multiple times and can thus be assumed to reflect the exact sequence of the donor genome. In contrast, $k$-mers which only appear once are assumed to contain sequencing errors. \cite{Melsted2011} and \cite{Marcais2011} make a more detailed, compelling argument about the importance of $k$-mer counting. 

In a genome of size $g$, we expect up to $g$ unique $k$-mers. This number can be smaller due to repeated regions (which produce the same $k$-mers) and small $k$, since smaller $k$-mers are less likely to be unique, but is usually close to $g$ for reasonable values of $k$. However, depending on the amount of sequencing errors, the total number of $k$-mers in the read library can be substantially larger than $g$. For example, in the DM dataset (Table \ref{data_sets}), the total number of 31-mers is approximately 289.20M while the number of 31-mers occurring at least twice is approximately 131.82M. The size of the genome is 122 Mbp (Mega base pairs). This is not surprising because one base call error in a read can introduce up to $k$ false $k$-mers. Consequently counting the frequency of all $k$-mers, as done  by Jellyfish~(\cite{Marcais2011}), which is limited to $k\leq 31$, requires $O(N)$ space where $N$ is the number of $k$-mers in the read library. This makes the problem of $k$-mer frequency counting intractable for large read libraries like Human even on a large machine with 256GB of memory. Ideally, the frequent $k$-mer identifier should use $O(n)$ space where $n$ is the number of frequent $k$-mers ($n\ll N$). The approach taken by BFCounter~(\cite{Melsted2011}) achieves something very close to this optimum by ignoring the infrequent $k$-mers with a Bloom filter~(\cite{Bloom1970}) and explicitly storing only frequent $k$-mers. This makes BFCounter much more memory efficient compared to Jellyfish. However, the running time of BFCounter is very large for two reasons. First, it is not multi-threaded. And second, both the Bloom filter and the Hash table used for counting incur frequent cache misses. The latter has recently been identified as a major obstacle to achieving high performance on modern architectures, motivating the development of cache-oblivious algorithms and data structures (\cite{Bender2005a}) which optimize the cache behavior without relying on information of cache layout and sizes. Additionally, BFCounter is also limited to a count range of 0-255 which will often be exceeded in single-cell experiments due to the large local coverage produced by whole genome amplification. A different approach is taken by DSK~(\cite{Rizk2013}) to improve memory efficiency. DSK makes many passes over the reads file and uses temporary disk space to trade off the memory requirement. Though \cite{Rizk2013} claimed DSK to be faster than BFCounter, on our machine using an 18TB Raid-6 storage system, DSK required more wall-clock time compared to BFCounter. Therefore, we consider DSK without dedicated high-performance disks like solid state and BFCounter to be too slow for practical use on large datasets.

We present a novel approach which reduces the memory footprint to accommodate large genomes and high coverage libraries. One of our tools (scTurtle) can report frequent 31-mers with counts (with a very low false positive rate) from a human read set with 146.5 Gbp using 109GB of memory in approximately 90 minutes using 19 threads. In contrast, Jellyfish did not complete the computation after 10 hours while consuming 185GB of memory, and BFCounter took 127 GB of memory but did not complete the computation after 64 hours. Like BFCounter, our approach also uses a Bloom filter to screen out $k$-mers with frequency one (with a small false positive rate), but in contrast to BFCounter, we use a Pattern-blocked Bloom Filter~(\cite{Putze2010}). The expected number of cache misses for each inquiry/update in such a Bloom filter is one. The frequency of the remaining $k$-mers are counted with a novel sorting and compaction based algorithm that we introduce in this article. Our compaction step is very similar to Run length encoding (\cite{data_comp}). Though the complexity of sorting is $O(n\log n)$, it has sequential and localized memory access which helps in avoiding cache misses and will run faster than an $O(n)$ algorithm that has $O(n)$ cache misses as long as $\log n$ is much smaller than the penalty issued by a cache miss. 

For larger datasets, where $O(n)$ space is not available, the method mentioned above will fail. We show that it is possible to get a reasonable approximate solution to this problem by accepting small false positive and false negative rates. The method is based on a counting Bloom filter implementation. The error rates can be made arbitrarily small by making the Bloom filter larger. Since the count is not maintained in this method, it only reports the $k$-mers seen more than once (with a small false positive and false negative rate), but not their frequency.

We call the first tool {\em scTurtle} and the second one {\em cTurtle}.

\section{Methods}
\subsection{scTurtle}
\subsubsection{Outline}
By a $k$-mer, we always refer to a $k$-mer and/or its reverse complement. Our objective is to separate the frequent $k$-mers from the infrequent ones and count the frequencies of the frequent $k$-mers. For that, first we use a Bloom filter to identify the $k$-mers that were seen at least twice (with a small false positive rate). To count the frequency of these $k$-mers, we use an array of items containing a $k$-mer and its count. These are the two main components of our tool. Once the counts are computed, we can output the $k$-mers having frequency greater than the chosen cutoff. For the sake of cache efficiency, the Bloom filter is implemented as a Pattern-blocked Bloom Filter~(\cite{Putze2010}). It localizes the bits set for an item to a few consecutive bytes (block) and thus reduces cache misses. The basic idea is as follows: when a $k$-mer is seen, the Bloom Filter is checked to decide if it  has been seen before. If that is the case, we store the $k$-mer in the array with a count of 1. When the number of items in the array cross a threshold, it is sorted in-place, and a linear pass is made, compressing items with the same $k$-mer (which lie in consecutive positions of the sorted array) to one item. The counts add up to reflect the total number of times a $k$-mer was seen. Note that, this strategy is very similar to run length encoding (\cite{data_comp}) the items. Our benchmarking (Table \ref{sac_hash}) shows that this very simple approach of storing items and their frequencies is faster than a Hash-table based implementation. An outline of the algorithm is given in Algorithm \ref{scTurtle_alg}. More details are provided in the following subsections. 


\begin{table}
\begin{center}
\caption{Comparison of Sort and Compress and Hash table based implementations for counting items and their frequencies. Jellyfish is a highly optimized hash table based implementation for the $k$-mer counting problem. We also compare against general purpose tools that uses Google sparse/dense Hash maps for storing $k$-mers and their counts.}
\label{sac_hash}
{\scriptsize

	\begin{tabularx}{86mm}{|>{\raggedleft}X|>{\raggedleft}X|>{\raggedleft}X|>{\raggedleft}X|r|}
	\hline
	\multirow{3}{*}{Method}& \multicolumn{4}{|c|}{Number of Insertions/Updates}   \\
	\cline{2-5}
	& \multicolumn{2}{|c|}{458M} &   \multicolumn{2}{|c|}{2.2B}  \\
	\cline{2-5}
		& Time(sec)& space(GB)& Time(sec)& space(GB) \\
	\hline
	\hline
	Sort and Compress &153.37 & 2.70 &   523.41 & 7.10 \\
	\hline
	Jellyfish & 296.49 & 2.40 &  1131.70 & 7.20 \\
	&  & &   &  \\
	\hline
	Google Dense Hash & 626.77 & 20.47 &  6187.95 & 40.38 \\
	\hline
	Google Sparse Hash & 1808.48 & 7.44 & 28069.18 & 10.60 \\
	\hline
	
	\end{tabularx}

}
\end{center}
\end{table}

\begin{algorithm}
\caption{scTurtle outline}
\label{scTurtle_alg}
\begin{algorithmic}[1]
\STATE Let, $S$ be the stream of $k$-mers coming from the read library, $BF$ be the Bloom filter, $A$ be the array to store $k$-mers with counts, $t$ be the threshold when we apply sorting and compaction. 
\FORALL {$k$-mer$\in S$}
	\IF {$k$-mer present in $BF$}
	\STATE Add $k$-mer to $A$
	\ENDIF
	\IF{$|A|\geq t$}
	\STATE Apply sorting and compaction on A
	\ENDIF
\ENDFOR

\STATE Apply sorting and compaction on A.
\STATE Report all $k$-mers in A with their counts as frequent $k$-mers and their counts.
\end{algorithmic}
\end{algorithm}

\subsubsection{$k$-mer extraction and bit-encoding}
For space efficiency, $k$-mers are stored in bit-encoded form where 2-bits represent a nucleotide. This is possible because $k$-mers are extracted out of reads by splitting them on `N's (ambiguous base calls) and hence contain only A, C, G and T. As we consider a $k$-mer and its reverse complement to be two representations of the same object, whenever we see a $k$-mer, we also compute the bit representation of the reverse complement and take the numerically smaller value as the unique representative of the $k$-mer/reverse complement pair.  

\subsubsection{Identification of frequent $k$-mers with Pattern-blocked Bloom Filter}
A Bloom filter is a space efficient probabilistic data structure which, given an item, can identify if this item was seen before with some prescribed, small false positive rate. We use this property of the Bloom filter to identify $k$-mers that were seen at least twice. An ordinary Bloom filter works as follows: A large bit-array($B$) of size $L$ is initialized to 0. Given an item $x$, $k$ hash values ($h_1, h_2, \ldots, h_k$) using $k$ independent hash functions (within the range $[0,(L-1)]$) are computed. We now check all the bits $B[h_1], \ldots, B[h_k]$. If they are all set to one, with high probability, this item has been seen at least once before. If not, it is certainly the first appearance of this item and we set all of $B[h_1], \ldots, B[h_k]$ to one. For all subsequent appearance(s) of this item, the Bloom filter will report that it has been seen at least once before. In this way, the Bloom filter helps us to identify frequent $k$-mers. Note that, if the bit locations are randomly distributed, due to the very large size of the Bloom filter, each bit inspection and update is likely to incur one cache miss. So, the total number of cache miss per item would be $k$. On the contrary, if the bit-locations are localized to a few consecutive bytes (a block), each item lookup/update will have a small number of cache misses. This can be done by restricting $h_1,\ldots,h_k$ to the range $[h_1(x), h_1(x)+b]$ where $b$ is a small integer. The bit pattern for each item can also be precomputed. This is called the Pattern-blocked Bloom Filter. \cite{Putze2010} observe that the increase in false positive rate due to this localization and precomputed patterns can be countered by increasing $L$ by a few percent. To summarize, we first select a block for an item (using a hash function), select $h_1<h_2< \ldots <h_k$ from a set of pre-computed random numbers such that all of them lie within the block and update/inquire them sequentially.

\subsubsection{Counting frequencies with sorting and compaction}
\begin{figure*}
\begin {center}
\caption{The Sorting and compaction mechanism. We start with an unsorted $k$-mer array. It is sorted and compacted (Phase-0 SAC). The empty part is filled with unsorted $k$-mers, sorted and compacted. After repeating this step several times, the compacted new part almost fills up the whole array. Then all items are sorted and compacted to produce Phase-1 SAC. This cycle repeats until all $k$-mers have been seen.}
\label{sort_comp_fig}
\includegraphics[scale=0.55]{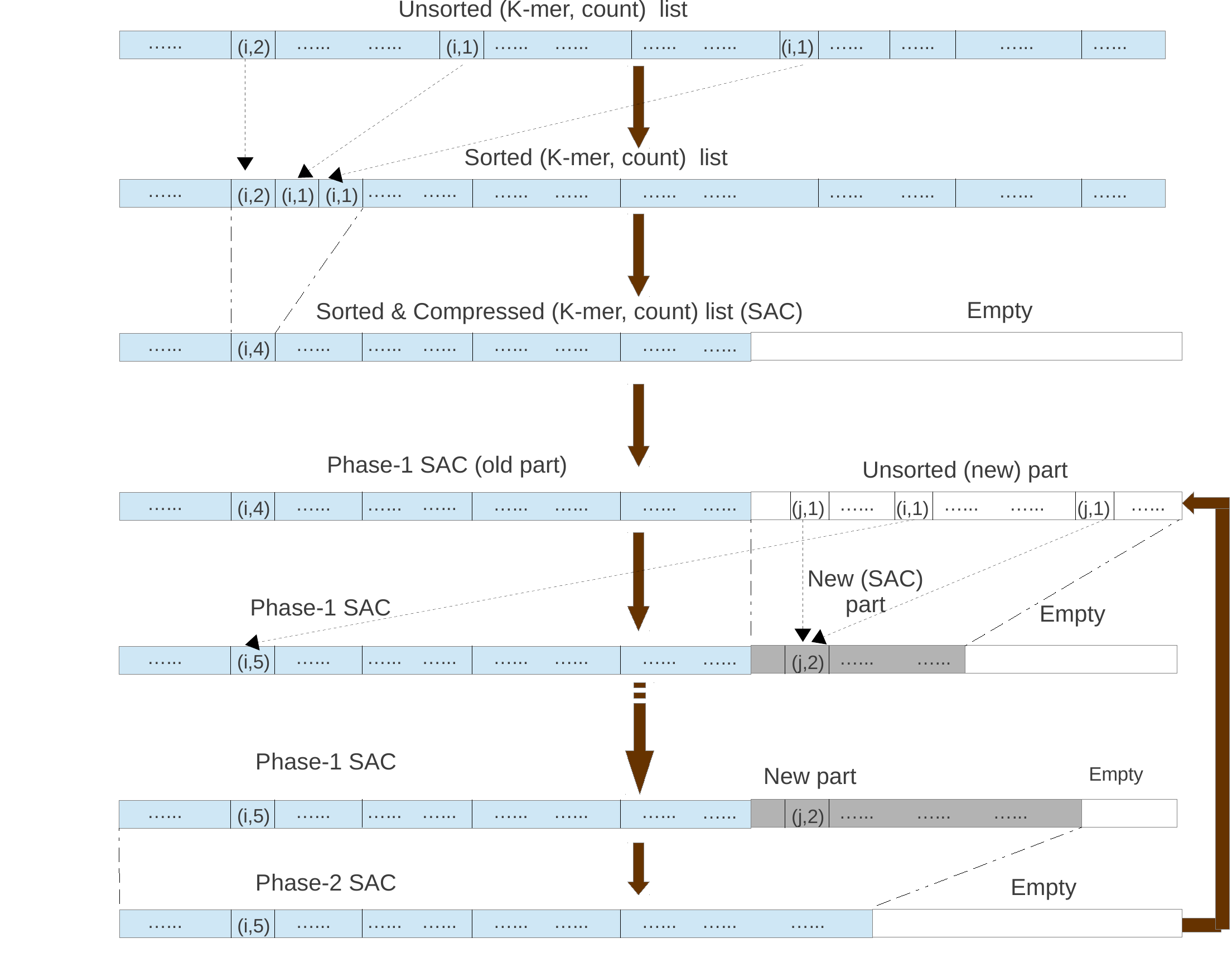}
\end {center}
\end{figure*}

Our next objective is to count the frequencies of the frequent $k$-mers. The basic idea is to store the frequent $k$-mers in an array $A$ of size $> n$, where $n$ is the number of frequent items. When this array fills up, we sort the items by the $k$-mer values. This places the items with the same $k$-mer next to each other in the array. Now, by making a linear traversal of the array, we can replace multiple items with the same $k$-mer with one item where a count field represents how many items were merged which is equal to how many times this $k$-mer was seen; see Figure \ref{sort_comp_fig}. Note that, this is very similar to run length encoding. Here is a toy example: Say $A=[\ldots, (i,1),\ldots, \ldots, (i,1), \ldots\ldots, (i,1)]$. After sorting $ A=[\ldots, \ldots, (i,1),(i,1),(i,1),\ldots \ldots]$ and compressing results in $A=[\ldots, (i,3), \ldots]$. We have to repeat these steps until we have seen all items. To reduce the number of times we sort the complete array, we apply the following strategy. We select a threshold $n<t<|A|$. We start with an unsorted $k$-mer array. It is sorted and compacted (Phase-0 Sorted and Compacted array or Phase-0 SAC). We progress in phases as follows. At phase $i$  a certain number of items in the beginning of the array are already sorted and compressed (Phase-($i-1$) SAC). The new incoming $k$-mers are stored as unsorted items in the empty part of the array. Let $m$ be the total number of items in the array. When $m>t$, we sort the unsorted items.  Many of these $k$-mers are expected to exist in Phase-($i-1$) SAC. We make a linear traversal of the array replacing $k$-mers present in both Phase-($i-1$) SAC and the newly sorted part with one item in Phase-($i-1$) SAC. $k$-mers not present in Phase-($i-1$) SAC are represented with one item in the newly sorted part. The counts are added up to reflect the total number of times a $k$-mer was seen. This takes $O(m)$ time. Note that this compaction has sequential and localized memory access which makes it cache efficient. After a few such compaction steps, $m>t$ and we sort and compress all the items in the array to produce Phase-$i$ SAC. 

By repeatedly applying this mechanism on the frequent items, we ultimately get the list of frequent $k$-mers with their counts decremented by 1. This is due to the fact that when inserted into the array for the first time, an item was seen at least twice unless its a false positive. To offset this, we simply add 1 to all counts before writing them out to a file. 

\setlength{\unitlength}{0.04cm}

\subsubsection{Parallelization}
We implemented a one producer, multiple consumer model with a pool of buffers. The producer extracts $k$-mers from the reads and distributes them among the consumers. Each consumer has its own Bloom filter. Since a $k$-mer should always pass through the same Bloom filter, we distribute the $k$-mers to the consumers using the modulo operation which is one of the cheapest hash functions available. Since modulo a prime number shows better hash properties compared to non-primes, it is recommended that one uses a prime (or at least an odd) number of threads as this spreads out the $k$-mers more evenly among the consumers which is helpful for speeding up the parallelization. $k$-mers are stored in buffers and only when the buffers fill up, they are transferred to the consumer. Since consumers consume the $k$-mers at an uneven rate, having the same fixed buffer size for all consumers may cause the producer to block if the buffer for a busy consumer fills up. To reduce such blocking, we have a pool of buffers that has more buffers than the number of consumers. If a consumer is taking longer to consume its items, the producer has extra buffers to store its $k$-mers in. This improves the speed-up. 

With many consumers (usually $>13$), the producer becomes the bottleneck. Therefore, its important to make the producer more efficient. The two most expensive parts of the producer are: converting reads to $k$-mers and the modulo operation required to determine which consumer handles a particular $k$-mer. Modern computers support SSE (\cite{comp_OandD}) instructions that operate on 128-bit registers and can parallely perform arithmetic/logic operations on multiple variables. We used SSE instructions for speeding up bit-encoding of $k$-mers. It is also possible to design approximate modulo functions that execute much faster than regular modulo instruction for some numbers (e.g. 5, 7, 9, 10, 63, etc) (\cite{Warren2012}). But each of these function have to be custom designed. If we restrict the number of consumers to the numbers that have efficient modulo function, its possible to improve the producer's running time even further. 

\subsubsection{Running time analysis}
We first analyze the sort and compress algorithm. Let the total number of frequent $k$-mers (those with frequency $\geq$ 2) be $N$ and let $n$ be the number of distinct frequent $k$-mers. We use an $xn, x> 1,$ sized array $A$ for storing the frequent $k$-mers and their counts. First consider the following simplified version of our algorithm: $(x-1)n$ new items are loaded into the array and they are sorted and compacted. Since there are $n$ distinct $k$-mers, at least $xn-n=(x-1)n$ locations will be empty after sorting and compaction. We again load $(x-1)n$ items and perform sorting and compaction. We iterate until all $N$ items have been seen. Each iteration takes $O(xn\log xn+ xn)$ time and we have at most $N/(x-1)n$ such iterations. So, the total time required is:
\begin{align*}
O\left(\frac{N}{(x-1)n}(xn\log xn+ xn)\right)=&O\left(\frac{x}{(x-1)}(N\log xn+ N)\right)\\
\leq & O\left(\frac{x}{(x-1)}(N\log N+ N)\right)
\end{align*}
As discussed earlier, to reduce number of times sorting is performed, which is much more expensive than compaction, we implemented a modified version of the above method which delays sorting at the expanse of more compactions. Our benchmarking shows this to be faster than the naive method. The algorithm we implemented progress in phases as follows. At the beginning of phase $i$, the array is filled up with unsorted elements. They are sorted and compacted ($O(xn\log xn+ xn)$). This is called the Phase-($i-1$) SAC. Let $e$ be the number of empty locations after each complete sorting and compaction step. Then, $(x-1)n\leq e \leq xn$. The new incoming $k$-mers are stored as unsorted items in the empty locations. When the empty part is full, we sort the new items ($O(xn\log xn)$). Many of these $k$-mers are expected to exist in Phase-($i-1$) SAC. We make a linear traversal of the array replacing $k$-mers present in both Phase-($i-1$) SAC and the newly sorted part with one item in Phase-($i-1$) SAC. $k$-mers not present in Phase-($i-1$) SAC are represented with one item in the newly sorted part. The counts are added up to reflect the total number of times a $k$-mer was seen. The total cost of a lazy compaction is therefore upper bounded by $O(xn\log xn+ xn)$. This again creates empty locations at the end of the array which allows us to perform another round of lazy compression. We assume that the incoming items are uniformly distributed and every lazy compaction stage reduces the size of the empty part by an approximately constant fraction $1/c$. Therefore, on average, we expect to have $c$ lazy compaction stages. This completes Phase-$i$, the expected cost of which is upper bounded by:  
\begin{align*}
&O(xn\log xn+ xn +c(xn\log xn+ xn))\\
=& O((c+1)(xn \log xn +xn))
\end{align*}
In order to compute how many phases are expected to consume all $N$ items, we observe that, at every phase, the lazy compaction steps consumes a total of at least $(x-1)n\{1+(1-\frac{1}{c})+(1-\frac{2}{c})+\ldots+ (1-\frac{c-1}{c})\}= (x-1)n(c+1)/2$ items. So, on average, each phase consumes at least $(c+1)n(x-1)/2$ items and hence, the expected number of phases is at most $2N/n(c+1)(x-1)$. Therefore, the total expected cost would be:
\begin{align*}
& \leq \frac{2N}{(c+1)n(x-1)}O\left( xn(c+1)\log xn +xn(c+1)\right)\\
&=\frac{2x}{(x-1)}O\left(N\log xn +N\right) \\
&\leq O\left(\frac{x}{(x-1)}(N\log N +N)\right) 
\end{align*}
Note that we obtained the same expression for the naive version of sorting and compaction. It is surprising that this expression is independent of $c$. As an intuitive explanation, observe that more lazy compactions within a phase results in more items being consumed by a phase, which in turn, decreases the number of phases. This inverse relationship between $c$ and the number of phases makes the running time independent of $c$. We found the naive version to be slower than the implemented version in empirical tests and therefore, believe our bound to be an acceptable approximation.

We now analyze the performance of sorting and compaction based strategy against a hash table based strategy for counting frequency of items. Let $p$ be the cache miss penalty, $h$ be the hashing cost, $s$ is the comparison and swapping cost for sort and compress and $b$ be the number of items that fit in the cache. The cost of frequency counting in the hash based method will be $(p+h)N$ since each hash update incurs one cache miss. For sorting and compress, we will have one cache miss for every $b$ operations and so, the cost for sorting and compaction will be $(p/b+s)a(N\log N+N)$, where $a=\frac{x}{(x-1)}$. To compute the value of $N$ for which sorting and compaction will be faster than a hash based method, we write:
\begin{align*}
(p+h)N \geq & (p/b+s)a(N\log N+N)\\
\log N \leq & \frac{(p+h)}{(p/b+s)a}-1
\end{align*}
Let a comparison and swap be one unit of work. A conservative set of values like $s=1, p=160$ (\cite{cache_penalty}), $h=8, b=256$ (assuming 8 bytes items and 2KB cache), $a=2$ results in $N \leq 2^{50}$. Therefore, for a large range of values of $N$, with a fast and moderate sized cache, the sorting and compaction based method would run faster than a hash based method.

Since every observed $k$-mer has to go through the Bloom filter, the time required in the Bloom filter is $O(M)$ where $M$ is the total number of $k$-mers in the read library. So, the total running time that includes the Bloom filter checks and sorting and compression of the frequent items is $O(M)+O(N\log N +N)$. Our measurements on the datasets used show that the total time is dominated by the Bloom filter updates (i.e. $O(M)>O(N\log N +N)$).

\subsection{cTurtle}
When there are so many frequent $k$-mers that keeping an explicit track of the $k$-mers and their counts is infeasible, we can obtain an approximate set of frequent $k$-mers by using a counting Bloom filter. Note that, the number of bits required for a Bloom filter for $n$ items is $O(n)$ but the constants are small. For example, it may be shown that for a 1\% false positive rate, the Bloom filter size is recommended to be approximately $9.6n$ bits (\cite{bloom_formula}). On the other hand, with a $k$-mer size of 32 and counter size of 1 byte, the memory required by any method that explicitly keeps track of the $k$-mers and their count is at least $9n$ bytes or $72n$ bits. So, the Bloom filter method will require much less memory. 

The basic idea of our counting Bloom filter is to set $k$ bits in the Bloom filter when we see an item for the first time. When seen for the second time, the item is identified as a frequent $k$-mer and written to disk. To record this writing, $k'$ more bits are set in the Bloom filter. For all subsequent sightings of this item, we find the $(k+k')$ bits set and know that this is a frequent $k$-mer that has already been recorded. For cache efficiency, we implement the counting Bloom filter as a Pattern-blocked counting Bloom filter as follows. We take a larger Bloom filter ($B$) of size $L$. When an item $x$ is seen, $k$ values ($h_1, h_2, \ldots, h_k$) within the range $[h(x),h(x)+b]$, where $h(x)$ is a hash function and $b$ is the block size, are computed using precomputed patterns. If this is the first appearance of $x$, with high probability, not all of the bits $B[h_1], \ldots, B[h_k]$ are set to one and so we set all of them. When we see the same item again, we will find all of $B[h_1], \ldots, B[h_k]$ set to one. We then compute another set of locations ($h_{k+1}, h_{k+2}, \ldots, h_{k+k'}$) within the range $[h(x)+b,h(x)+2b]$ using precomputed patterns. Again, with high probability, not all of $B[h_{k+1}], \ldots, B[h_{k+k'}]$ are set to 1 and so we set all of them. At the same time we write this $k$-mer to disk as a frequent $k$-mer. For all subsequent observations of this $k$-mer, we will find all of $B[h_1], \ldots, B[h_{k+k'}]$ set to 1 and will avoid writing it to disk. Note that a false positive in the second stage means that we don't write the $k$-mer out to file and thus have a false negative. 

Currently, cTurtle reports $k$-mers with frequency greater than one. But this strategy can be easily adopted to report $k$-mers of frequency greater than $c>1$. We argue that for most libraries with reasonable uniform coverage, $c=1$ is sufficient. Let $C$ be the average nucleotide coverage of a read library with read length $R$. Then the average $k$-mer coverage is $C_k=\frac{C(R-k+1)}{R}$~(\cite{Zerbino2008}). Suppose we have an erroneous $k$-mer with one error. The probability that the same error will be reproduced is $\frac{1}{3k}$ where $1/k$ is the probability of choosing the same position and 1/3 is the probability of making the same base call error. Therefore, the expected frequency of that erroneous $k$-mer is $1+\frac{C_k-1}{3k}$. For $R=100, k=31$, this expression is $1+0.0075C$. So, we need $C>132.85$ at a location for an erroneous 31-mer to have frequency greater than 2. Since most large libraries are sequenced at much lower depth ($<60x$),  such high coverage is unlikely except for exactly repeated regions and therefore, our choice of frequency cutoff will provide a reasonable set of reliable $k$-mers. However, this does not hold for Single Cell libraries which exhibit very uneven coverage (\cite{Chitsaz2011}), but note that, frequent $k$-mers are considered reliable only for uniform coverage libraries and thus single cell libraries are excluded from our consideration.

The  parallelization strategy is the same as that for scTurtle.

\section {Comparisons with existing $k$-mer counters}
The datasets we use to benchmark our methods are presented in Table \ref{data_sets}. The library sizes range from 3.7 Gbp to 146.5 Gbp for genomes ranging from 122Mbp to 3.3Gbp. To the best of our knowledge, currently Jellyfish~(\cite{Marcais2011}) is the fastest and DSK~(\cite{Rizk2013}) is the most memory efficient open-source $k$-mer counters. BFCounter~(\cite{Melsted2011}) uses an approach similar to ours and is memory efficient but in contrast to our method, not computationally efficient. 

\begin{figure}
\caption{CPU utilization curve for scTurtle (top) and cTurtle (bottom). The diagonal shows the theoretical optimum. The deviation from the optimum is largely due to the bottleneck of having a single threaded producer for extracting and distributing $k$-mers.}
\label{cpu_util}
\includegraphics[scale=0.45] {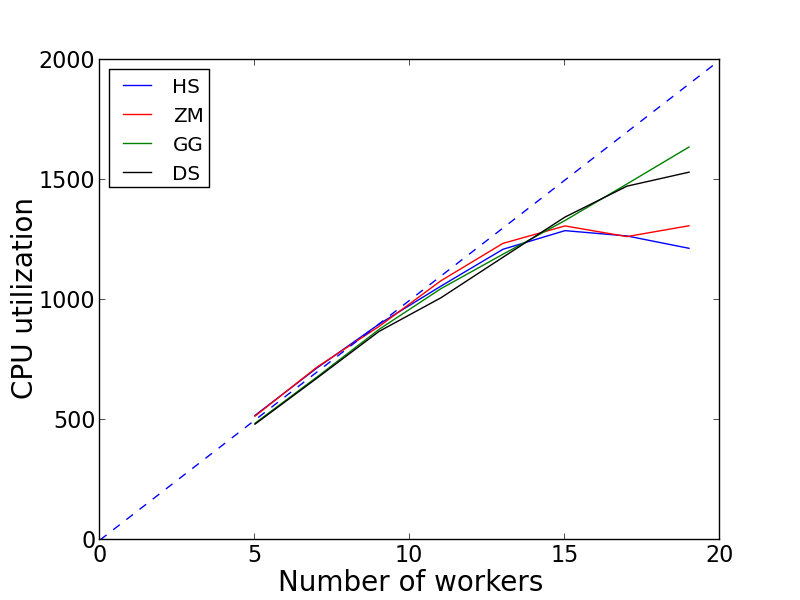}
\includegraphics[scale=0.45] {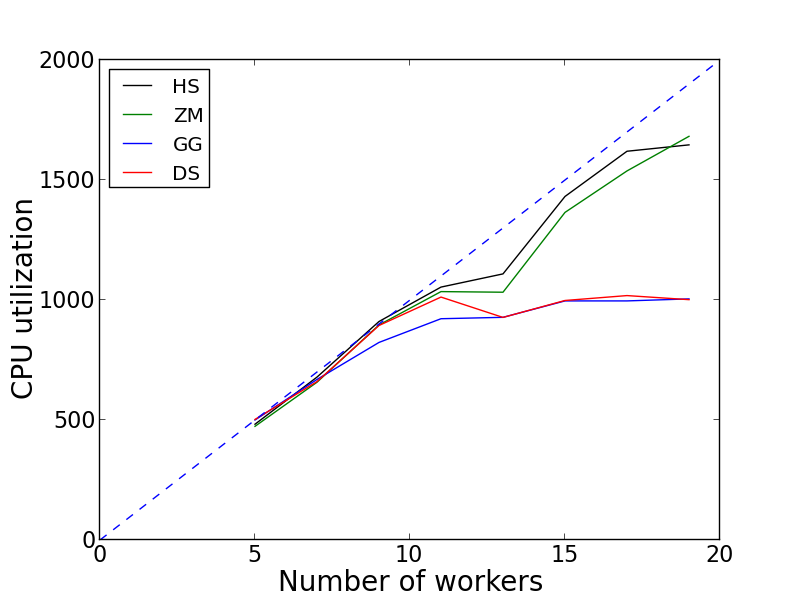}
\end{figure}

On the two small datasets, Jellyfish runs faster but uses more memory than scTurtle ($1.7\times$ faster using $1.71\times$ memory for GG dataset with 19 threads). Jellyfish's lower wall-clock time is mainly due to its parallelization scheme, which is not limited to one producer for generating and distributing the $k$-mers---the major bottleneck for the Turtles (Figure \ref{cpu_util}). On the two large datasets, Jellyfish did not produce results after running for more than 10 hours on a machine with 256GB memory, while our tools could work with less than 128GB. Unexpectedly, on these large datasets, BFCounter required more memory than scTurtle (over $128$GB vs. $109$GB). We suspect this is due to the memory overhead required to reduce collusions in the hash table which we avoid using our sort and compaction algorithm. \cite{Rizk2013} claimed DSK to be faster than BFCounter, but on our machine, which had a 18TB Raid-6 storage system of 2TB SATA disks, it proved to be slower (1591 mins vs. 1012 mins for the GG dataset). \cite{Rizk2013} reported using more efficient storage systems (like SSD), the lack of which in our machine might explain DSK's poor performance in our experiments. The detailed results are presented in Table \ref{comparative_results} for multi-threaded Jellyfish, scTurtle and cTurtle. Since BFCounter (single threaded) and DSK (4-threads) do not allow variable number of threads, we present their results separately in Table \ref{bfc_dsk}.

To validate our claim that the wall-clock time (and therefore parallelization) may be improved by speeding up the producer, we made special versions of scTurtle and cTurtle with are 31-threads and a fast approximate modulus-31 function. For the largest library tested (HS), on average, the special version of scTurtle produces frequent 31-mers in approximately 73 minutes compared to approximately 90 minutes by the regular version (a 19\% speedup). As we use 64-bit integers for storing $k$-mers of length 32 and less and 128-bit integers for storing $k$-mers of length in the range 33 to 64, the memory requirement for larger $k$-mers were also investigated. Again for the largest dataset tested (HS), we found that scTurtle's memory requirement increased from 109GB for $0< k\leq 31$ to 172GB for $32\leq k \leq 64$ (a 58\% increase). Note that the Turtles require less memory for up to 64-mers than Jellyfish for 31-mers.  Detailed results of all the datasets for the Turtles are presented in Table \ref{large_kmer_time}. 

We also examined the error rates for our tools and BFCounter. Note that, just like BFCounter, scTurtle has false positives only and cTurtle has both false positives and false negatives. We investigated these rates for the two small datasets (see Table \ref{false_rates}) and found error rates for all tools to be smaller than 1\%. For the large datasets, due to memory requirements, we could not get exact counts for all $k$-mers and therefore could not compute these rates. 


\begin{table}
\begin{center}
\caption{Descriptive statistics about the datasets used for benchmarking. The library sizes range from 3.7Gbp to 146.5Gbp and the genome size ranges from 122Mbp to 3.3Gbp.}
\label{data_sets}
{\scriptsize
	\begin{tabular}{|c|c|r|c|r|}
	
	\hline
	Set ID& Organism & Genome Size (Mbp) & Read Lib & Bases (Gbp)  \\
	\hline
	\hline
	DM &  {\it D. Melanogaster} & $122$ &SRX040485 & 3.7 \\
	\hline
	GG & {\it G. Gallus} & $1\times 10^3$ & SRX043656 & 34.7\\
	\hline
	ZM & {\it Z. Mays} & $2.9\times 10^3$ & SRX118541 & 95.8\\
	\hline
	HS & {\it H. Sapiens} & $3.3\times 10^3$ & ERA015743 & 146.5\\
	\hline
	\end{tabular}
}
\end{center}
\end{table}

\begin{table*}
\begin{center}
\caption{Comparative results of Wall clock time and memory between scTurtle, cTurtle and Jellyfish. Each reported number is an average of 5 runs. The $k$-mer size is 31. Recall that scTurtle and Jellyfish report $k$-mers and their counts, while Porkmer only reports the $k$-mers with count $>1$.}
\label{comparative_results}
  \begin{tabular}{|l|l|r|r|r|r|r|r|r|r|r|r|}
  \hline
	\multirow{2}{*}{Set ID} & Tool & \multicolumn{8}{|c|}{Multi-threaded Wall clock time (min:sec)}	& Memory\\
  \cline{3-10}
				& 	& 5 & 7 & 9 & 11    & 13   & 15   &  17  & 19 &  (GB) \\
   \hline
   \hline
   \multirow{3}{*}{DM}   
   &	 Jellyfish 		&3:59  &  2:59   & 2:24    & 1:59  & 1:42 &  1:22&   1:13 & 1:12 &  7.4 \\
   \cline{2-11}
   &	 scTurtle		    & 4:55  &   3:37  &   2:48 &  2:34 &  2:28 &  2:30 &  2:24 & 2:20 & 5.5 \\
	\cline{2-11}
   &	 cTurtle 	    &  3:41 &   2:43  &    2:04  & 1:55 &  1:55 &  1:55 &   1:55 & 1:56 & 4.2   \\
   \hline  
   \hline  
  \multirow{3}{*}{GG}
   &	 Jellyfish 		& 46:23  &  34:11  & 27:32  & 22:46  & 19:44 & 17:29  & 15:38  & 14:19 &  81.9 \\
   \cline{2-11}
   &	 scTurtle		    &  56:42 & 40:20 & 33:24 & 30:07  & 28:06 &  25:57 &  25:16 & 25:52 & 47.1 \\
	\cline{2-11}
   &	 cTurtle 	    & 45:16  &   33:04  &  23:40  & 21:37 & 21:06 &  21:16 &  21:34 & 21:21 &   29.9 \\
   \hline
   \hline  
  \multirow{3}{*}{ZM}
   &	 Jellyfish 		& N/A  &  N/A  & N/A    & N/A  & N/A & N/A  & N/A  & N/A &  N/A \\
   \cline{2-11}
   &	 scTurtle		    & 171:36  &   114:24  &  95:00 & 77:48 & 68:48 &  65:09 &  67:24 & 67:12 & 82.1\\
	\cline{2-11}
   &	 cTurtle 	    & 131:00  &   98:48  &   72:48    & 65:12 &  65:36 &  65:12 &  63:24 & 63:48 &  51.6  \\
   \hline
   \hline  
   \multirow{3}{*}{HS}
   &	 Jellyfish 		& N/A  &  N/A  & N/A    & N/A  & N/A & N/A  & N/A  & N/A &  N/A \\
   \cline{2-11}
   &	 scTurtle		    &   212:00 &  152:36 &  124:36 & 106:24 & 97:12 &  89:36 &  89:12 & 89:36 & 109.5 \\
	\cline{2-11}
   &	 cTurtle 	    &  171:00 &  123:12  &   98:00   & 87:12 & 91:12  &  89:24  & 88:00  & 90:24 & 68.5   \\
   \hline
  \end{tabular}
\end{center}
\end{table*}

\begin{table}
\begin{center}
\caption{Performance of BFCounter and DSK (4 threads) for 31-mers. Some of the results are not available since those computations could not be completed within a reasonable time.}
\label{bfc_dsk}
{\scriptsize
  \begin{tabularx}{86mm}{|>{\raggedleft}X|>{\raggedleft}X|>{\raggedleft}X|>{\raggedleft}X|r|}
  \hline
 Set ID & Tool & Wall-clock time (min:sec)& CPU Utilization (\%) & Space (GB)\\
  \hline			
  \hline
   \multirow{2}{*}{DM}  
     &	 BFCounter 		& 78:35 & 99 & 3.24\\
   \cline{2-5}
    &	 DSK  & 170:37 & 318 & 4.86\\
  \hline
   \multirow{2}{*}{GG}  
     &	 BFCounter 		& 1011:51 & 99 & 29.26\\
   \cline{2-5}
    &	 DSK   & 1590:54 &  290 & 48.59\\
  \hline
   \multirow{2}{*}{ZM} 
     &	 BFCounter 	& $>$2289:00 & NA & $>$166.00\\
   \cline{2-5}
    &	 DSK  &  $>$2923:00 & NA & NA\\
  \hline
    \multirow{2}{*}{HS}   
    &	 BFCounter 		& $>$3840:00 & NA & $>$128.00 \\
   \cline{2-5}
    &	 DSK  & $>$1367:00 & NA & NA\\
   \hline
  \end{tabularx}
}
\end{center}
\end{table}

\begin{table}
\begin{center}
\caption{Performance of scTurtle and cTurtle for 64-mers. The tools ran with fast mod and 31 threads. Each reported number is an average of 5 runs.}
\label{large_kmer_time}
{\scriptsize
  \begin{tabularx}{86mm}{|>{\raggedleft}X|>{\raggedleft}X|>{\raggedleft}X|>{\raggedleft}X|>{\raggedleft}X|r|}
  \hline
	$k$-mer size & Set ID & Tool & Wall-clock time (min:sec)& CPU Utilization (\%)& Space (GB)\\
  \hline			
  \hline
  \multirow{8}{*}{31}    
   &\multirow{2}{*}{DM}   &	 scTurtle 		& 02:18 & 2335.8 & 5.50\\
   \cline{3-6}
   & &	 cTurtle    & 1:28 &  580.0 & 4.20 \\
  \cline{2-6}
     
   & \multirow{2}{*}{GG} &	 scTurtle 		& 24:48 & 2388.0 & 47.10\\
   \cline{3-6}
   & &	 cTurtle    & 28:51 &  413.2 & 29.90 \\
  \cline{2-6}
    
   & \multirow{2}{*}{ZM}  &	 scTurtle 		& 55:57 & 1838.0 & 82.15\\
   \cline{3-6}
   & &	 cTurtle    & 74:46  &  756.0 & 51.60 \\
  \cline{2-6}
   
   & \multirow{2}{*}{HS}   &	 scTurtle 		& 73:24 & 1563.0 & 109.53\\
   \cline{3-6}
   & &	 cTurtle    & 98:24  &  512.0 & 68.55\\
   \hline
   \hline
   \multirow{8}{*}{48}    
   &\multirow{2}{*}{DM}   &	 scTurtle 		& 2:33 & 1790.0 & 8.30\\
   \cline{3-6}
   & &	 cTurtle   	&2:30 & 871.4 & 4.76 \\
  \cline{2-6}
     
   & \multirow{2}{*}{GG} &	 scTurtle 			& 25:11 &  1373.8 & 70.69\\
   \cline{3-6}
   & &	 cTurtle    	&25:29  & 693.8 & 29.34\\
  \cline{2-6}
    
   & \multirow{2}{*}{ZM}  &	 scTurtle 			& 90:16 &  1125.0 & 129.09\\
   \cline{3-6}
   & &	 cTurtle    	& 81:28 &  782.0 &52.35 \\
  \cline{2-6}
   
   & \multirow{2}{*}{HS}   &	 scTurtle 			&112:11 & 953.4 &172.11\\
   \cline{3-6}
   & &	 cTurtle    	&105:10 &  657.6&69.29 \\
   \hline
   \hline
  \multirow{8}{*}{64}    
   &\multirow{2}{*}{DM}   &	 scTurtle 		& 1:40 & 948.0 & 8.30\\
   \cline{3-6}
   & &	 cTurtle    & 1:28 &  580.0 & 4.76 \\
  \cline{2-6}
     
   & \multirow{2}{*}{GG} &	 scTurtle 		& 31:60 & 825.8 & 70.69\\
   \cline{3-6}
   & &	 cTurtle    & 28:51 &  413.2 & 29.35 \\
  \cline{2-6}
    
   & \multirow{2}{*}{ZM}  &	 scTurtle 		& 79:90 & 1037.0 & 129.09\\
   \cline{3-6}
   & &	 cTurtle    & 74:46  &  756.0 & 52.35 \\
  \cline{2-6}
   
   & \multirow{2}{*}{HS}   &	 scTurtle 		& 79:56 & 952.0 & 172.11\\
   \cline{3-6}
   & &	 cTurtle    & 98:24  &  512.0 & 69.29 \\
   \hline
  \end{tabularx}
}
\end{center}
\end{table}

\begin{table}
\begin{center}
\caption{False Positive and False Negative rates of scTurtle and cTurtle. For the large datasets, due to memory constraints, the exact counts for all $k$-mers could not be obtained and therefore these rates could not be computed.}
\label{false_rates}
{\scriptsize
\begin{tabular}{|c|c|c|c|c|}
\hline
\multirow{2}{*}{Set ID}& scTurtle & BFCounter &\multicolumn{2}{|c|}{cTurtle}   \\
\cline{4-5}
	& (FP only)(\%) & (FP only)(\%) & FP (\%) & FN (\%) \\
\hline
\hline
DM &  0.003	&0.300 & $1.9\times 10^{-4}$	& $2.3\times 10^{-4}$	\\
\hline
GG & 0.848	& 0.027 & 0.31	& 0.08	\\
\hline
\end{tabular}
}
\end{center}
\end{table}

\section{Conclusion}
Identifying correct $k$-mers out of the $k$-mer spectrum of a read library is an important step in many methods in bioinformatics. Usually, this distinction is made by the frequency of the $k$-mers. Fast tools for counting $k$-mer frequencies exist but, for large read libraries, they may demand a huge amount of memory which can make the problem unsolvable on machines with moderate memory resource ($\leq 128$ GB). Simple memory efficient methods, on the other hand, can be very time consuming. Unfortunately there is no single tool that achieves a reasonable compromise between memory and time. Here we present a set of tools that make some compromises and simultaneously achieves memory and time requirements that are matching the current state of the art in both aspects. 

In our first tool (scTurtle), we achieve memory efficiency by filtering $k$-mers of frequency one with a Bloom Filter. Our Pattern-blocked Bloom filter implementation is more time-efficient compared to a regular Bloom filter. We present a novel strategy based on sorting and compaction for storing frequent $k$-mers and their counts. Due to its sequential memory access pattern, our algorithm is cache efficient and achieves good running time. However, because of the Bloom filters, we incur a small false positive rate.

The second tool (cTurtle) is designed to be more memory efficient at the cost of giving up the frequency values and allowing both false positives and false negatives. The implementation is based on a counting Bloom filter that keeps track of whether a $k$-mer was observed and whether it has been stored in external media or not. This tool does not report the frequency count of the $k$-mers.

Both tools allow $k$-mer size of upto 64. They also allow the user to decide how much memory should be consumed. Of course, there is a minimum memory requirement for each dataset and the amount of memory directly influences the running time and error rate, but we believe, with the proper compromises, the frequent $k$-mer extraction problem is now approximately solvable for large read libraries within reasonable wall-clock time using moderate amount of memory.

\bibliographystyle{abbrv}
\bibliography{my_bib}

\end{document}